\documentstyle[prc,aps,eqsecnum,epsf,amssymb]{revtex}

\newcommand{\bmr}{{\mbox{\boldmath $r$}}}

\newcommand{\bmq}{{\mbox{\boldmath $q$}}}
\newcommand{\bmb}{{\mbox{\boldmath $b$}}}

\newcommand{\bmR}{{\mbox{\boldmath $R$}}}

\newcommand{\qq}{{|\bmq|}}

\newcommand{\jac}{\xi}
\newcommand{\jacb}{{\mbox{\boldmath $\xi$}}}
\newcommand{\hypfi}{\varphi}
\newcommand{\Fz}{\phi_0}
\newcommand{\Fu}{\phi_1}
\newcommand{\Fn}{\phi_n}

\begin{document}
\preprint {WIS-01/04 Oct-DPP}
\draft
\title{ GRS  computation of deep inelastic electron 
scattering on $^4$He}
\author{M. Viviani and A. Kievsky }
\address{INFN, Sezione Pisa, and Phys. Dept., University of
Pisa, I-56100 Pisa, Italy}
\author{A.S. Rinat}
\address{Department of Particle Physics, Weizmann Institute of Science,
         Rehovot 76100, Israel}
\maketitle
\begin{abstract}
We compute cross sections for inclusive scattering of high energy
electrons on  ${}^4$He, based on the
two lowest orders of the Gersch-Rodriguez-Smith (GRS) series.
The required one- and two-particle density matrices
are obtained from non-relativistic ${}^4$He wave 
functions using realistic models for
the nucleon-nucleon and three-nucleon interaction.  Predictions for
$E$=3.6 GeV agree well with the NE3 SLAC-Virginia data.
\end{abstract}

\section{Introduction}
\label{sec:intro}

Total nuclear structure functions (SF) contain at least two
components, namely one describing nuclei  as composed of
point-particles and a second one  accounting for the internal
structure of the nucleons.  Since the latter are taken from experiment, a 
computation of total nuclear SF  amounts to a determination of the same 
for a nucleus which is composed of
point-particles. Such a calculation is usually performed  within one of the 
following two approaches. 
In the first, one  perturbatively expands  the SF in the residual 
interaction  between a nucleon struck by the virtual photon
and the remaining spectator nucleus,  thus generating the Impulse Series (IS) 
for the SF. In the widely used lowest order Impulse Approximation  
this residual  interaction is first neglected. One then 
either computes  higher order Final State Interaction (FSI) terms (see
for instance Refs. \cite{nik,rj}), or models them \cite{ciof1,omar,cor,ciof3}. 

An alternative approach is based on a 
relativistic generalization of the Gersch-Rodriguez-Smith (GRS) 
expansion of SF in inverse powers of the 3-momentum transfer 
$\qq$ \cite{grs,sag1,gr1}. Both theories have been applied to
cross sections for inclusive scattering of high-energy
leptons from  various nuclear targets \cite{ciof1,omar,cor,ciof3,rt1,rtt}.

When  applied to high energy inclusive scattering one usually limits
a GRS calculation to the two lowest order terms. Their determination 
requires knowledge of one- and two-particle density
matrices, which are  not diagonal in  the coordinate  of the
struck  nucleon  $'$1$'$, and of the spectral function. The
non-diagonal one-body density  
matrix  is related to the single-nucleon momentum distribution
$n(p)$ and  is usually extracted 
from alternative experimental sources,  or is computed from theoretical 
models.  There generally is no direct information
on the half-diagonal, two-particle  density matrices for finite 
systems and  one  relies on 
parametrizations~\cite{grs,rt2}. In those, nuclear recoil is usually 
neglected, thereby limiting applications to targets with $A\gtrsim 12$.

In the following we exploit accurately computed non-relativistic (NR)
wave functions for light nuclei, using a number of modern realistic
nucleon-nucleon ($NN$) and three-nucleon ($3N$) interactions. Those wave
functions are  
Galilean invariant and enable a  realistic GRS calculation 
of inclusive scattering  on those nuclei. As a first application we 
choose $^4$He and for that target we  shall report below predictions and
a comparison with data. 

The other ingredient, namely the $^4$He spectral function ${\cal
P}(p,{\cal E})$, is rather difficult to compute and only a few
direct calculations are reported~\cite{MS91,efrosS}. 
Below we shall adopt the reasonable alternative that has been
developed by Ciofi degli Atti and Simula~\cite{ciof3}.

At this point we  mention that for years the IS and GRS approaches have 
been considered as being distinct and even incompatible.
Only recently has their equivalence been demonstrated, provided 
both series are expanded to the same order in the same parameter 
\cite{rj,sag1,gr1}. Following the derived
prescription to link the two approaches, one can 
perform an interesting numerical comparison.  

The GRS and IS theories are not the only tools which have been used to
compute  nuclear SF. We mention in particular the ingenious method
of Efros and co-workers, which has been applied to $^4$He \cite{efrosS,efros}. 
Regrettably, it appears not feasible to extend that method to high energies.

The present note is organized as follows.
In section~\ref{sec:cal} we recall the GRS approach,
emphasizing the two main ingredients of our calculations, namely the
SF of a target composed of point-particles and the SF of the free
nucleons. We also discuss  there the computation of the above density  
matrices. In  section~\ref{sec:res} we compare  predictions for cross sections
with the Virginia-SLAC data~\cite{ne3}. 
In the last section we present
our conclusions.

\section {Total nuclear structure functions.}
\label{sec:cal}

The cross section per nucleon for inclusive scattering of 
high-energy electrons from a nucleus with $A$ nucleons reads
\begin{eqnarray} 
  \frac{d^2\sigma_{eA}(E;\theta,\nu)/A}{d\Omega\,d\nu}
  =
  \frac{2}{M}\sigma_M(E;\theta,\nu)\bigg\lbrack\frac {xM^2}{Q^2}
  F_2^A(x,Q^2)+ {\rm tan}^2(\theta/2)F_1^A(x,Q^2) \bigg\rbrack\ ,
\label{a1}
\end{eqnarray}
where $M$ is the nucleon mass, $\sigma_M$ the Mott cross section,
$E$ the beam energy, $\theta$ the laboratory  scattering
angle and $\nu$ the energy loss imparted onto the target. The  above  nuclear 
structure functions  per nucleon $F_k^A(x,Q^2)$ contain the essence of 
unpolarized  electron scattering from  randomly  oriented  targets. Those SF
depend on  the squared 4-momentum transfer $-q^2=Q^2=\bmq^2-\nu^2$ and
on the Bjorken variable  $x=Q^2/2M\nu$ 
with range $0\le x\le A$. For given  beam energy $E,\,\,\,(\theta,\nu)$
and $(x,Q^2)$  are sets of alternative   kinematic variables.

Total nuclear structure functions per nucleon may, in a semi-heuristic
fashion, be expressed  as follows~\cite{gr2,rt3,commar} 
\begin{eqnarray}
  F^A_k(x,Q^2)=\int_x^A\frac {dz}{z^{2-k}}f^{PN,A}(z,Q^2)
  \sum_{l=1}^2
  C_{kl}({Q^2\over x^2},z)
  F_l^{\langle N\rangle}\bigg (\frac{x}{z},Q^2\bigg )   \ ,
\label{a2}
\end{eqnarray}
where  $f^{PN,A}$ is the SF of a nucleus
composed  of point-particles and $F_k^{\langle N\rangle}$ is
the averaged free nucleon SF. For  a nucleus $A(Z,N)$   
\begin{equation}
   F_k^{\langle N\rangle}(x,Q^2)={Z\over A} F^p_k(x,Q^2)+
   {N\over A} F^n_k(x,Q^2)\ , \qquad k=1,2                  \ .
\end{equation}
The coefficient functions $C_{kl}$ account for the 
mixing and modification  of the free nucleon structure
functions in the expression (\ref{a2})~\cite{atw}. We 
retained in this paper only the dominant coefficient  
\begin{eqnarray}
   C_{22}({Q^2\over x^2},z)&\approx& [1-\rho+\rho/z]^2-{1\over 2}
   \rho(1-\rho)(1-1/z)^2\ ,  \nonumber\\
\noalign{\medskip}
   \rho&=&[1+Q^2/4M^2x^2]^{-1}\ .
\label{a97}
\end{eqnarray} 
Eq. (\ref{a97}) is a better approximation for $C_{22}$ than
previously used~\cite{commar}.

Eq.~(\ref{a2}) is valid for $x\gtrsim 0.15-0.20$, below which pionic and 
anti-screening effects become of importance \cite{pion} and above some
critical $Q_c^2$, which presumably can be estimated using QCD.
A previous comparison of  predictions and data for medium-$A$ targets 
produced an empirical estimate  $Q^2\gtrsim Q_c^2\approx 2.0-2.5\,$GeV$^2$
~\cite{rtt}.

Each of the  SF $F^{p,n}_k$ in (\ref{a2})
has both nucleon-elastic (NE) and nucleon-inelastic (NI) parts, thus
$F_k^{N}=F_k^{N,{\rm NE}}+F_k^{N,{\rm NI}}$, with $N=p,n$. 
The total nuclear structure functions, Eq.~(\ref{a2}), and the total cross 
section per nucleon may therefore be expressed as a
sum over contributions coming from the NE and NI parts of nucleon SF.
In particular, the  NE part $F_k^{N,{\rm NE}}$ is the well-known
combinations of  static 
electro-magnetic  form factors and contributes primarily 
around the region of the quasi-elastic  peak (QEP),
$x\approx 1$.   For the inelastic parts $F_k^{n,{\rm NI}}$ we have
taken 
\begin{equation}
     F^{n,{\rm NI}}_k(x,Q^2)\approx F_k^{d,{\rm NI}}(x,Q^2)-
     F_k^{p,{\rm NI}}(x,Q^2)\ ; \qquad k=1,2\ ,
\end{equation}
where $F^{d,{\rm NI}}_k(x,Q^2)$ are the deuteron SF's per nucleon. 
For  $F^{p,{\rm NI}}_1(x,Q^2)$ and  
$F^{d,{\rm NI}}_1(x,Q^2)$ we employ values interpolated between the data of 
Ref.~\onlinecite{bod}, whereas for $F^{p,{\rm NI}}_2(x,Q^2)$ 
and  $F^{d,{\rm NI}}_2(x,Q^2)$ we use the parametrizations of
Ref.~\cite{amad}.

\subsection{The GRS series} \label{sec:grs}

We now focus on $f^{PN,A}$ in Eq.~(\ref{a2}), the SF for a nucleus of 
point-particles, which has to be computed.
Following Ref.~\cite{rt1} one writes
\begin{equation}\label{eq:f2phi}
 f^{PN,A}(x,Q^2)= \left( \frac{\partial y_G}{\partial x}\right)
  _{Q^2\ {\rm fixed }}
  \phi(\qq,y_G)\ ,\qquad\qq=Q\sqrt{1+\bigl(Q/2Mx\bigr)^2}\ ,
\end{equation}
where $\phi(\qq,y)$ is the  reduced response
in terms of a relativistic scaling variable  \cite{sag1} 
\begin{eqnarray}
y_G=y_G^{\Delta}&\approx &y_G^{\infty}\bigg [1-\frac{1}{2A'}\frac{\nu^2}{\qq^2}\xi+
{\cal O}(1/A'^2)\bigg ]\ , \qquad A'=A-1\ ,
\nonumber\\ 
y_G^{\infty}&=&\frac{M\nu}{\qq}\xi\ ,
\nonumber\\ 
\xi&=&\bigg(1-\frac{\langle \Delta\rangle}{M}-x\bigg )\ ,
\label{a4}
\end{eqnarray}
and $\langle \Delta \rangle$ some average nucleon separation
energy. We shall
retain the above $1/A'$ correction  in the scaling variable $y_G$  which, 
as Eq. (\ref{a4}) shows, is simply related to the Bjorken variable 
$x$.  In the GRS approach the reduced response may, for 
smooth $NN$
interactions,  be expanded  in a series of  inverse powers of
$\qq$~\cite{grs,gr1}. Explicitly  
\begin{eqnarray}
   \phi(\qq,y_G)= \Fz(\qq,y_G)+
     \sum_{n\ge 1}
       \bigg (\frac {M}{\qq}\bigg )^{n}  \Fn(y_G)\ .
\label{a6}
\end{eqnarray}
The lowest order term is given by
\begin{eqnarray}
  \Fz(\qq,y_G^{\Delta_0})=
   \int_{|y_G^{\Delta_0}|}^{\infty} \frac{dp\;
  p}{4\pi^2}\int_0^{E_M} 
   d{\cal E}\;{\cal P}(p,{\cal
E})+\theta(y_G^{\Delta_0})\int_0^{y_G^{\Delta_0}} 
  \frac{dp\; p}{4\pi^2}\int_{E_m}^{E_M} d{\cal E}\;{\cal P}(p,{\cal
  E})\ ,
  \label{aa6}
\end{eqnarray}
with ${\cal P}(p,{\cal E})$ the standard single-hole spectral
function. The  energy argument is ${\cal E}=E-\Delta_0$ with $E$ the removal
energy and $\Delta_0$ the ($p,n$ averaged)
minimal separation energy (for $^4$He $\Delta_0\approx 20.2$
MeV). Above, $y_G^{\Delta_0}$ is the  scaling variable   
give in Eq. (\ref{a4}) with $\langle\Delta\rangle=\Delta_0$.
The integration limits in (\ref{aa6}) are
\begin{eqnarray}
E_{M\atop m}(y_G,p,\qq)=\frac{(y_G \pm p)\qq}{\nu}\ .
\label{aa7}
\end{eqnarray}
In actual calculations the spectral function  has been written
as in ~Ref.~\cite{ciof4} 
\begin{equation}
  {\cal P}(p,{\cal E})= n_0(p)\delta({\cal E})+{\cal P}_1(p,{\cal E})\ ,
  \label{eq:sf}
\end{equation}
where $n_0(p)$ is the partial momentum distribution due to
intermediate states of one nucleon and the $A-1$ spectator system
in its ground state. Contributions from continuum states of that
system are summed in ${\cal P}_1(p,{\cal E})$. As stated in the
Introduction,  that part of the spectral function for $^4$He has been
taken to be ${\cal P}_1(p,{\cal E})= 
{\cal N}(p) {\cal 
P}^{\rm model}_1(p,{\cal E})$, where ${\cal P}^{\rm model}_1(p,{\cal
E})$ has been  provided by us by C. Ciofi degli Atti~\cite{ciof3}. The
normalization factor ${\cal N}(p)$  is fixed by
\begin{equation}
  \int_{{\cal E}_{\rm thr}}^{\infty} d{\cal E}\;{\cal P}_1(p,{\cal E})=
  n(p)-n_0(p) \ ,
  \label{eq:sfno}
\end{equation}
where the quantities $n(p)$, the total momentum distribution,
and $n_0(p)$ have been calculated using the NR wave functions 
as  will be explained in Subsect.~\ref{sec:chh}. 

We have also tested the
following simple  2-state  approximation for the 
spectral function \cite{ciof4,efros}
\begin{eqnarray}
  {\cal P}(p,{\cal E})\approx n_0(p)\delta({\cal E})+
     [n(p)-n_0(p)] \delta({\cal E}-\langle \Delta \rangle+\Delta_0)\ ,
   \qquad\langle \Delta \rangle\approx 50\ {\rm MeV}\ .
  \label{a7}
\end{eqnarray}
Substitution into (\ref{aa6}) produces a $\qq$-independent
lowest order contribution,
\begin{equation}
    \Fz^{(1)}(y_G^\Delta,y_G^{\Delta_0})=
     \frac{1}{4\pi^2}\bigg \lbrack \int_{|y_G^{\Delta}|}
      ^{\infty} dp\,p\, n(p)-
        \int_{|y_G^{ \Delta}|}^{|y_G^{\Delta_0}|}
    dp\,p\,n_0(p)\bigg \rbrack\ . 
  \label{aa8a}
\end{equation}
Since in the relevant $p$ region $n_0\approx n$, 
an accurate approximation of Eq.~(\ref{aa8a}) reads
\begin{equation}
     \Fz^{(2)}(y_G^{\Delta_0})  
      \approx \frac{1}{4\pi^2}
       \int_{|y_G^{\Delta_0}|}^{\infty} dp\,p\, n(p) \ .
  \label{aa8b}                                         
\end{equation}

Terms with $n \ge 1$ in Eq. (\ref{a6}) describe
FSI corrections to the asymptotic limit as a 
series in $1/\qq$. It is easier to give  those in terms of their
Fourier transform $\tilde \Fn(s)$, 
namely
\begin{equation}\label{eq:phiqs}
   \Fn(y_G^\Delta)=\int_0^\infty {ds\over 2\pi} \; e^{isy_G^\Delta}\;
   \tilde \Fn(s)\ . 
\end{equation}
Each   $\tilde \Fn(s)\equiv \tilde
\Fn(s;[V])$ is a functional of the bare interaction $V$, for instance 
\begin{mathletters}
\label{a85}
\begin{eqnarray}
   \frac{M}{\qq}\tilde \Fu(s)& =& \int d\bmr_1\int d\bmr_2\;
   \rho_2(\bmr_1,\bmr_2;\bmr_1-s\hat\bmq,\bmr_2)[i\tilde\chi_q(\bmb,z;s)]
   \ ,
\label{a85a}\\
  \tilde\chi_q(\bmb,z;s)&=&\tilde\chi_q^{(1)}(\bmb,z;s)+
   \tilde\chi_q^{(2)}(\bmb,z;s)  \ ,
\label{a85b}\\
 \tilde \chi_q^{(1)}(\bmb,z;s)&=& -\frac{M}{\qq}\int_0^s d\sigma
   V(\bmb,z-\sigma)\ , 
\label{a85c}\\
 \tilde\chi_q^{(2)}(\bmb,z;s)&=&
     \frac{M}{\qq}sV(\bmb,z-s) =
   -s {\partial\over\partial s}\tilde \chi_q^{(1)}(\bmb,z;s) \ ,    
\label{a85d}
\end{eqnarray}
\end{mathletters}
where $\bmb$ ($z$) is the component of the vector $\bmr=\bmr_1-\bmr_2$
perpendicular (parallel) to the  $\bmq$ direction,  and $\rho_2$ the
semi-diagonal two-particle density matrix.  Eqs. (\ref{a85}) define
two parts of the off-shell eikonal phase $\tilde\chi_q$ which are 
related, and thus
\begin{equation}
\tilde \chi_q(\bmb,z;s)= \bigg (1-s{\partial\over\partial  s}\bigg )
\tilde \chi_q^{(1)}(\bmb,z,s)\ .
\label{a86}
\end{equation}
One frequently deals with interactions $V$ which have
a strong short-range repulsion (or produce for other reasons a diffractive 
elastic amplitude) and it is then of advantage to perform a summation 
over a ladder of bare interactions $V$. The replacement $V\to
V_{eff}= t_q$, produces a well-behaved, $q$-dependent, off-shell
$t$-matrix as an effective interaction, which in coordinate space
is proportional to the off-shell profile function $\tilde\Gamma$ \cite{rt1}.  
For the
part $\tilde\Gamma^{(1)}$, generated by $\tilde\chi^{(1)}$, one has
\begin{mathletters}
 \label{a87}
\begin{eqnarray}
  i\tilde\chi_q^{(1)}(\bmb,z;s) &\to& 
  \tilde \Gamma^{(1)}_q(\bmb,z;s)=
   {\rm   exp}[i\tilde\chi_q^{(1)}(\bmb,z;s)]-1 \ , \label{a87a}\\
   &\approx&\theta(z)\theta(s-z)\Gamma_q^{(1)}(b)\ .
\label{a87b}
\end{eqnarray}
\end{mathletters}
The approximation Eq. (\ref{a87b}) has been tested in Ref. 
\cite{rt4}.  Its application permits the exploitation of a standard 
parametrization of the on-shell profile $\Gamma^{(1)}_q(b)$  in terms
of elastic   $NN$ scattering data,
as are $\sigma_q^{tot}, \tau_q, Q^{(0)}_q$, which are  
respectively the total cross section, the ratio of the real to imaginary part
of the forward elastic amplitude and the width of the diffractive
amplitude. Explicitly,
\begin{equation}\label{a70}
   \Gamma_q^{(1)}(b) \approx \frac{1}{2}\sigma_q^{tot}(1-i\tau_q)\;
   {[Q^{(0)}_q]^2\over 4\pi} \;e^{-[bQ^{(0)}_q]^2/4}\ .
\end{equation}

There is no simple way to generalize (\ref{a86}) to the total
off-shell phase $\tilde\chi$.  Yet as in \cite{rt1}
we shall assume that  Eq.~(\ref{a86})  is also approximately valid
for the total off-shell profile function 
\begin{mathletters}
 \label{a87bis}
\begin{eqnarray}
  i\tilde\chi_q(\bmb,z;s) \to 
  \tilde \Gamma_q(\bmb,z;s)&\approx&
   \bigg \lbrack 1-s{\partial\over\partial s}\bigg \rbrack
    \tilde \Gamma_q^{(1)}(\bmb,z;s) \ , \label{a87bisa} \\
    &\approx&
   \bigg \lbrack 1-s{\partial\over\partial s}\bigg \rbrack \theta(z)\theta(s-z)
       \Gamma_q^{(1)}(b)\ . \label{a87bisb}
\end{eqnarray}
\end{mathletters}
After substitution of the above expression in Eq.~(\ref{a85}), the
leading FSI contribution to $\tilde\phi(\bmq,s)$ turns
into the following $q$-dependent result  ~\cite{rt1} 
\begin{equation}
   {M\over \qq}\tilde \Fu(s;[V]) \to \frac {M}{\qq}\tilde \Fu(s,[t])=
\tilde G_1(\qq,s)
   =\tilde G_1^{(1)}(\qq,s)+\tilde G_1^{(2)}(\qq,s)\ ,
  \label{a88}
\end{equation}
where
\begin{mathletters}
\label{a8}
\begin{eqnarray}
   \tilde G_1^{(1)}(\qq,s) &\approx&-\int d\bmr_1\int d\bmr_2\;
   \rho_2(\bmr_1,\bmr_2;\bmr_1-s\hat\bmq,\bmr_2)
    \theta(z)\theta(s-z)\Gamma^{(1)}_q(b) \ ,
   \label{a8a}\\
     \tilde G_1^{(2)}(\qq,s)&\approx&\int d\bmr_1 \int d\bmr_2\;
    \rho_2(\bmr_1,\bmr_2;\bmr_1-s\hat\bmq,\bmr_2)
    s\theta(z)\delta(s-z) \Gamma^{(1)}_q(b)\ .
\label{a8b}
\end{eqnarray}
\end{mathletters}
Previous analyses dealt with targets with $A\ge 12$. For those there 
do not exist computations from first principles for  single-nucleon  
momentum distributions $n_0(p)$, $n(p)$ and density matrices $\rho_2$,
as required in (\ref{eq:sf}), (\ref{eq:sfno}), (\ref{a8}). 
Moreover, suggested parametrizations \cite{grs,rt2} do not   
account for  nucleon-recoil, which is only justified for $A\gtrsim$ 12. 

One of the predictions from Eq.~({\ref{a2}) is a  weak-$A$  dependence of 
the SF for  point-nucleon nuclei and of the averaged nucleon SF
\cite{rt1}.  This entails predicted inclusive cross sections per nucleon
to be practically independent of $A$.
Supporting evidence comes from experimental ratios of
cross sections per nucleon for different targets  at identical
kinematical conditions~\cite{rt1,rtt,arr1}. 
Definitely  larger deviations from  smooth  $A$-dependence are expected, 
if  one of the targets is a light nucleus  with $A\le$6. This is
evident from Table~\ref{tab:rat}  where we entered some 
C/Fe ratios from JLab data  ~\cite{arr} and for He/C 
from the older NE3 data Ref.~\cite{ne3}.

For the above reasons we did not include in the past
a GRS analysis of inclusive scattering on the lightest targets.
In the following we exploit the possibility to compute a precise  NR nuclear
ground state wave function $\Phi_0$ of light nuclei for given nuclear
interaction  $V$. Those enable  a calculation  of
$n(p),n_0(p)$ and $\rho_2$, which enter the components 
(\ref{eq:sf}), (\ref{eq:sfno}) and~({\ref{a8}) of the nuclear SF.

\subsection{The density matrices}
\label{sec:chh}

Various methods permit nowadays an accurate calculation of the ${}^4$He ground 
state wave function~\cite{kam}.
We exploit here the  Correlated Hyperspherical Harmonic function 
(CHH) technique which has been developed by the Pisa group. 
The spatial configuration of the system is described in terms of a
given choice of the Jacobi vectors $\jacb_1,\jacb_2,\jacb_3$.
In the hyperspherical framework we use as new variables
the hyperradius $\rho$, defined by
\begin{equation}\label{eq:rho}
         \rho^2=\sum_{i=1}^3 \jac_i^2\ ,
\end{equation}
and  the set $\Omega=\{\hat \jac_1,\hat\jac_2,\hat
\jac_3,\hypfi_2,\hypfi_3\}$. The latter includes the polar
angles $\hat \jac_i\equiv (\theta_i,\phi_i)$  of each Jacobi
vector and  additional   hyperspherical 
angles $\hypfi_2$, $\hypfi_3$.
We then write for the ground state wave function $\Phi_4$ 
\begin{equation}
   \Phi_4 = \sum_{n=1}^{N_{tot}} 
   \Bigg [ {u_{n}(\rho) \over \rho^4}
    {\cal A} \biggl\{ {\cal F}_n(r_{12},r_{13},r_{14},r_{23},r_{24},r_{34})\;
    {\cal Y}_{n}(\Omega) \biggr\}
   \Bigg ] \ , 
\label{a100}
\end{equation}
where ${\cal A}$ is an anti-symmetrizer and $ {\cal
Y}_{n}(\Omega)$ are the four-body Hyperspherical 
Harmonic (HH) functions~\cite{F83}. Choosing ${\cal F}_n=1$,
Eq.~(\ref{a100}) generates an uncorrelated  HH expansion for the 
${}^4$He ground state wave function. For it, the rate of convergence 
is  extremely slow when the $NN$ interaction is strongly repulsive 
at small distances. 
One accounts for the latter property by
multiplying every HH function  in the expansion with a suitably 
chosen correlation factor ${\cal F}_n$, ultimately   leading to the
CHH expansion. 
The latter much improves the description of
the target wave function for small inter-nucleon distances, and a much 
smaller number of basis functions is required to get convergence. 

In the  case of $^4$He, the correlation  factors have been chosen to be of the
Jastrow form~\cite{KRV95}
\begin{equation}
    {\cal F}_n 
      = f_{n}(r_{12}) g_{n}(r_{13}) g_{n}(r_{14})
         g_{n}(r_{23}) g_{n}(r_{24}) g_{n}(r_{34}) \ ,
     \label{eq:corre4}
\end{equation}
i.e. products of one-dimensional functions
$f_{n}$ and  $g_{n}$ which  are solutions of a $NN$
Schr\"odinger-like  equation (for  details, see
Ref.~\cite{KRV95}). 

Using the Rayleigh-Ritz variational principle  for varying
functions $u_n$ in (\ref{a100})
\begin{equation}
<\delta_u\Phi_4|H-E|\Phi_4>=0 \ ,
\label{eq:rrvar}
\end{equation}
one is led  to a set of hyperradial equations for the functions $u_n$
in the variable $\rho$ which, after discretization,  is
converted into a
generalized eigenvalue problem and are solved by standard numerical
techniques \cite{KRV94}. One thus determines
the hyperradial functions $u_{n}(\rho)$ in
Eq.~(\ref{a100}) and the bound state energy $E$. 
We shall present below results  based on 

1) the Argonne V18 nucleon-nucleon (NN) potential~\cite{AV18} (the
   AV18 model),  

2) the Argonne V18 NN  potential supplemented by 
the Urbana IX three--nucleon potential~\cite{PPCPW97}
(the AV18UR model), 

3) the Argonne V14 NN potential~\cite{AV14}
plus the Urbana VIII three--nucleon  potential~\cite{W91} (the AV14UR
model). 

The two models, which  contain a  three-nucleon interaction   provide
a ${}^4$He 
binding energy rather close to the experimental one, whereas the AV18
under-binds  by  about 4 MeV.
The present status of the  $^4$He binding energy calculations with the
CHH method is summarized in Table~\ref{tab:be},
where in Eq.~(\ref{a100})  up to  $N_{tot}\approx 200$
functions have been used (the explicit CHH states included in the
expansion are  discussed in Ref.~\cite{KRV95}).

Calculated binding energies for the AV18 or AV18UR
Hamiltonians are within 1 \% of the ``exact" Green's function Monte-Carlo 
(GFMC) results \cite{PPCPW97} for corresponding interactions. 
Somewhat less satisfactory agreement between the CHH and GFMC
results for the AV14UR model, since this interaction is more
repulsive at short distances than the other two. For all we 
checked that our final results for the deep inelastic scattering cross 
sections depend only  slightly on the value of $N_{tot}$, once
$N_{tot}\gtrsim200$. 

The thus constructed ground state wave function, readily gives
the corresponding
4-body density matrices, in particular the one non-diagonal in $'1'$
\begin{equation}
  \rho_4(\bmr_1,\bmr_2,\bmr_3,\bmr_4;\bmr_1',\bmr_2,\bmr_3,\bmr_4)=
  \Phi_4(\bmr_1,\bmr_2,\bmr_3,\bmr_4)\Phi_4(\bmr_1',\bmr_2,\bmr_3,\bmr_4)
\end{equation}
Successive integrations  over the diagonal coordinates 3,4,  
and eventually over coordinate 2, then furnish $\rho_2$ and
$\rho_1$ non-diagonal in $'1'$.  The total momentum distribution
is the Fourier transform of $\rho_1$.

The partial momentum distribution $n_0(p)$ is obtained from the
overlap of $\Phi_4$ and the (${}^3$H, ${}^3$He averaged)
three-nucleon
ground state wave function $\Phi_3$, namely
\begin{equation}
  n_0(p)=|a(p)|^2\ ,\qquad
  a(p)=\langle\; \Phi_3(1,2,3)\chi_4\eta_4 | j_0(p R_{123,4})
  \Phi_4(1,2,3,4)\; \rangle\ ,\label{eq:n0}
\end{equation}
where $\chi_4$ ($\eta_4$) is the spin (isospin) state of particle 4,
$j_0$ the zero-order Bessel function and $R_{123,4}$ the distance of
particle 4  with respect to the center of mass of the other
three. The ground state  wave function $\Phi_3$ of the three-nucleon
system  has been obtained  with the same Hamiltonian model used to
generate $\Phi_4$, and again by application of the CHH
technique~\cite{KRV94}. In equation~(\ref{eq:n0}), $\Phi_3$ and
$\chi_4\eta_4$ are coupled to give a state with vanishing total
angular momentum and isospin.

Eqs. (\ref{a8}) for $\tilde G$ invite to express $\rho_2$
in terms of the  variables $\bmr=\bmr_1-\bmr_2$ and
$\bmR=(\bmr_1+\bmr_2)/2$  and then to perform
the $\bmR$ integration 
\begin{equation}
  B(\bmb,z;s)=\int d\bmR\; 
      \rho_2(\bmr_1,\bmr_2;\bmr_1-s\hat\bmq,\bmr_2)
\label{a81}
\end{equation}
The result is substituted into (\ref{a8}) and subsequently
integrated  over  $\bmb$ and $z$.

\subsection{An effective IS series}
\label{sec:is}

In Ref. \onlinecite{ciof1} one may find a detailed account
of the lowest order PWIA  calculations as part of the IS series. To it one
should perturbatively  add FSI due to the
interaction of the knocked-out nucleon and the spectator core.

In the Introduction we  recalled
a recent proof~\cite{gr1,rj} that  the GRS and IS, both evaluated  to order 
${\cal O}(1/\qq^2)$, produce  the same result.  The lowest order is
given by 
\begin{eqnarray}
  \Fz(\qq,y_0)=
   \int_{|y_0|}^{2|{\mbox{\boldmath $q$}}|+y_0} \frac{dp\;
  p}{4\pi^2}\int_0^{{\bar E}_M} 
   d{\cal E}\;{\cal P}(p,{\cal
   E})+\theta(y_0)\int_0^{y_0} 
  \frac{dp\; p}{4\pi^2}\int_{{\bar E}_m}^{{\bar E}_M} d{\cal E}\;{\cal P}(p,{\cal
  E})\ ,
  \label{aa6is}
\end{eqnarray}
where in this case~\cite{ciof1,gr1}
\begin{equation}
  {\bar E}_{M\atop m}=\sqrt{M^2+(y_0+\qq)^2}-\sqrt{M^2+(p\mp \qq)^2}\ .
  \label{eq:emm}
\end{equation}
Above, $y_0$ is the IS scaling variable defined by  \cite{ciof1,gr1}
\begin{eqnarray}
   y_0&=&y_0^{\infty}(1+\delta_0)\ , \nonumber\\
   y_0^{\infty}&=&-\qq+\sqrt{(\nu-\Delta_0)(\nu-
   \Delta_0 +2M)}\ , \nonumber\\
   \delta_0&\approx&-\frac{1}{2A'}\bigg \lbrack \frac{1+\nu/M}
   {1+\qq/y_0^{\infty}} \bigg \rbrack + {\cal O}(1/A'^2)\ .
\label{a5}
\end{eqnarray}
The IS first order is then obtained  from  the prescription which puts   
$\tilde G_1^{(2)}=0$ in (\ref{a88}) and simultaneously replaces 
the GRS scaling variable $y_G^{\Delta}$, Eq. (\ref{a4}),  by $y_0$
in Eq.~(\ref{eq:phiqs}).
Finally, $f^{PN,A}(x,Q^2)$ is now calculated using
again $y_0$ in place of $y_G^\Delta$ in Eq.~(\ref{eq:f2phi}).
Those changes have been shown to generate the IS
series to  order $1/\qq$ \cite{rj}. Actual applications will 
be given in the following section.

\section{Results}
\label{sec:res}

In this section we  present  numerical  results for 
inclusive cross sections on $^4$He based on Eqs. (\ref{a1}), (\ref{a2}). 
Unless stated differently, those results  use density  matrices 
based on  the  AV18UR  interaction  and the full spectral function
as discussed in Sec.~\ref{sec:grs}.  In the GRS approach,
the cross sections have been computed by approximating  the reduced
structure function as follows
\begin{eqnarray}
    \phi(\qq,y_G) & \to  \Fz(\qq,y_G^{\Delta_0})\ ,
     \phantom{+ G_1(\qq,y_G^{\Delta_0})
     \equiv \phi_{01}(\qq,y_G^\Delta,y_G^{\Delta_0})}
     & \qquad{\rm 0\ order}\ ,\\
    \phi(\qq,y_G) & \to 
     \Fz(\qq,y_G^{\Delta_0})+ G_1(\qq,y_G^{\Delta_0})
     \equiv \phi_{01}(\qq,y_G^{\Delta_0})\ ,
      & \qquad{\rm 0+1\
   order}\ , 
\end{eqnarray}
where $\Fz$ is given by Eq.~(\ref{aa6}) and $G_1$ is the inverse Fourier 
transform of the functions given in Eq.~(\ref{a8}).  The corresponding IS 
expressions are obtained as discussed in Sect.~\ref{sec:is}.

In  Fig.~1  we display the SLAC-Virginia
NE3 cross section data~\cite{ne3} for $E$=3.595 GeV,  scattering angles
$\theta=$16$^\circ$, 20$^\circ$, 25$^\circ$, 30$^\circ$ and varying energy 
loss $\nu$, and also our numerical results 
obtained with the above reduced response $\phi_{01}$.  
Table \ref{tab:ne3} shows the ranges of the kinematical variables 
$x,\nu,Q^2$. For all  measured scattering angle,  and in addition 
for $\theta=45^{\circ}$, we entered there values of the energy loss 
$\nu$ and of $Q^2$ close to the appropriate lower ($x=2$) and upper
($x=0.1$) ends of the theoretical curves shown in Fig. 1. We also show 
the $\nu,Q^2$ values corresponding to the position of the QEP at $x=1$ .

Let us first focus  on the NE part of the cross section (thin
dashed lines), which contribute primarily around the QEP. 
As can be seen from the figure, the NE parts well describe the QEP,
in particular at low $Q^2$.  The data for 
$\theta=16^{\circ}$ show a clear  maximum and an adjacent minimum which get
fuzzier and ultimately disappear for increasing $\theta$ or $Q^2$. The same 
maximum occurs in the inclusive cross sections on 
deuterons \cite{arrd}, but is absent for targets with $A\ge 12$ \cite{ne3,arr}.
 
This structure is the result of
the competition between the NE and the NI parts 
of the cross section. For decreasing $Q^2$  the NE part around the QEP $x=1$ 
grows relative to the, usually dominant, NI part.
For the smallest $Q^2$ in the data, the NE part beyond the QEP 
stands out until for $x\lesssim 1$ the NI one overtakes. 
It is the maximum value of $f^{PN,A}(x,Q^2)$ 
(for $x\approx 1$) which sets the magnitude of the QEP.
For a given $Q^2$, that peak value decreases with $A$: it is largest
for the deuteron and  $^4$He and then it is almost independent on $A$
for $A\ge 12$,  reflecting the smearing of the momentum distribution
due to the Fermi motion. For example,  $f^{PN,A}(x=1,Q^2)$ for the
deuteron  and $^4$He is $\approx 5.5$ and $2.2$ times, respectively,
larger than for nuclei with $A\ge 12$.  It causes the QEP for equal
kinematic conditions to be most prominent for the lightest nuclei.

We already mentioned that Eq.~(\ref{a2}) (for the NI part) is
estimated to be valid for $x\gtrsim 0.2$ and above some critical 
$Q_c^2\approx 2.0-2.5$ GeV$^2$~\cite{rt1,rtt}. As Table~\ref{tab:ne3} shows, 
that approximate  critical value is actually never  reached for any 
$\theta =16^{\circ}, 20^\circ$ data point,
which  renders those
data  not really suitable for a test of the theory. For the same reason we 
excluded from our analysis NE3 data at lower energies and the same is 
the case for old, near-elastic, high-$Q^2$ data on $^4$He \cite{rock}.
For both angles, the convolution formula predicts too large NI
parts around the QEP, resulting in an over prediction of the data in
that region. For $\theta=30^\circ$, on the other hand, 
$Q^2\gtrsim Q_c^2$ and, in fact, comparison  of data and
predictions shows that there is good  agreement for all but the
smallest energy loss values. Since  cross sections there   
have fallen  by orders of magnitude, one expects sensitivity to 
small dynamical details.  For example, without the inclusion in 
Eq. (\ref{a2}) of the mixing factor $C_{22}$, Eq. (\ref{a97}), the agreement 
would  be of definitely lower quality  (see also below). 

A more stringent test for the theory would be provided by data at
higher  beam energies with, in general, higher $Q^2$. Unfortunately,
the recent 4 GeV experiments at JLab for various targets did not 
contain $^4$He~\cite{arr}, but a recent JLab proposal includes that
target  in a 6 GeV run with scattering angles
$\theta=15^\circ$, $23^\circ$, $30^\circ$, $45^\circ$ and $60^\circ$
\cite{arr2}. The kinematical region explored by that 
experiment covers $0.2<x<1.0$ and $1<Q^2<8.0$ GeV$^2$
(Table~\ref{tab:arr})  and predictions  for the four largest 
scattering angles can be found in Fig.~2. Incidentally, we checked
that for $E=6$ GeV the effect of $C_{22}$ is practically 
negligible due to  $Q^2$, which grows with beam energy 
$E$.

Next we discuss the effect of the different approximations for
$\Fz$ presented in Sec.~\ref{sec:cal}. The cross sections calculated
using the approximations~(\ref{aa8a}) and~(\ref{aa8b}) for $\Fz$ 
almost overlap. Moreover,  they are rather similar to 
the ones computed  using the spectral function, Eq.~(\ref{aa6}),
except in the low $\nu$ region.  There, the use of the  ``full''
model slightly reduces the cross sections.

This result deserves some comment. The differences
between the two expressions, Eqs.~(\ref{aa6}) and~(\ref{aa8b}) 
are generally sizeable in particular in the low $\nu$
region~\cite{ciof1} where $y_G^{\Delta_0}$ is negative and large in absolute
value. However, in the kinematical region of the NE3 experiment, the
values of $E_M$  entering Eq.~(\ref{aa6}) are found to be rather large
and then 
\begin{equation}
  \int_0^{E_M} d{\cal E} {\cal P}(p,{\cal  E})
   \approx   \int_0^\infty d{\cal E} {\cal P}(p,{\cal  E})
   =n(p)\ .\label{eq:limit}
\end{equation}
As a result, the $\Fz$ calculated using  Eqs.~(\ref{aa6}) or~(\ref{aa8b})
nearly coincide. For example, for $E=3.595$ GeV, $\theta=30^\circ$ and 
$\nu=0.7$ GeV, $\qq/\nu\approx 3$ and $E_M$ is large for all the
values of $p$.  
For the same reason we expect that the  predicted cross sections do not much 
depend on the parametrization chosen for the spectral function, in particular 
not in the low $\nu$ region. In that region 
they  are rather sensitive to the tail of the momentum
distributions, which in turn is  related to the correlations in
the nuclear wave functions~\cite{lyka}. 

In Fig. 3a we display the separate contributions of the cross sections 
at $E=3.595$ GeV, $\theta=30^\circ$ and varying $\nu$, as due to
the NE and NI components of the nucleon SF  $F_k^{\langle N\rangle}$.
The thin (heavy) dashes are  NE part in
the $\phi_0$ ($\phi_{01}$) approximation for 
the reduced response. 
Those have their maximum at $x\approx 1$, 
and are only in the wings  marginally affected by the 1st order FSI terms. 
The thin and heavy  solid  lines show the corresponding
NI parts, which by nature  dominate the region
$x\lesssim 1$ for relatively high  $\nu$.  FSI affect only 
the low $\nu$ region and cause  a rather small increase in  cross
sections. In those $\nu$-regions NE and NI parts are of the
same order. Fig. 3b is as Fig.~3a  for $\theta=45^\circ$  (this
angle was chosen since the $Q^2$ values are larger and similar to
those for $E=6$ GeV, $\theta=15^\circ$).
The  results are similar, except that  FSI now appear to
decrease the 0th NI contribution at low $\nu$. 

From Fig.~3a we observe that the slight
over-prediction in the low $\nu$ region of the theoretical results at
$\theta=30^\circ$  is mainly due to the NI part and as
discussed above, is only marginally affected by the inclusion of  FSI. 
The reason of the large NI contribution can be
simply understood by looking at the convolution~(\ref{a2}) between the
nucleon SF and $f^{PN,A}$.

Typical  behavior  of the functions $f^{PN,A}(z,Q^2)$,
$F^{\langle N\rangle,NI}_2(x/z,Q^2)$ and the mixing factor
$C_{22}(Q^2/x^2,z)$, (the latter two for $x=0.1, 2.0$) are given in 
Fig.~4 ($F^{\langle N\rangle,NI}_1$ behaves similarly). 
For  $x<1$, the permitted 
range of values $z\ge x$ covers the  $z\approx 1$ region where
$f^{PN,A}(z,Q^2)$ is large and allows virtually the entire support $x/z$ of
$F^{\langle N\rangle, NI}$ to contribute.
In contrast, for $x>1$ (see Fig.~4b),  only the
tail of $f^{PN,A}$ contributes and $F^{\langle N\rangle,NI}(t,Q^2)$ for
$x/A\le t\le1$ is usually small. Moreover, $f^{PN,A}$ decreases
as $z\rightarrow 4$. In fact, as $z$ becomes larger, 
also $|y_G^{\Delta_0}|$ increases and the integral in Eq.~(\ref{aa6})
decreases. However, $y_G^{\Delta_0}\approx -{\sqrt{ Q^2}}/2$ for $z\gg
1$ and since the 
values of $Q^2$ in the $E=3.595$ GeV NE3 experiment are not large, $\Fz$ and
the corresponding $f^{PN,A}$ are still non-vanishing at $z\rightarrow 4$. As
a consequence, the integrals receive a sizeable contribution from the
region $z>3$ and the NI cross section in the low $\nu$ region remains
large.

Note also that the mixing factor $C_{22}\approx 1$ for $x=0.1$
(Fig.~4a) but becomes rather small for $x=2$ (Fig.~4b),
sizeably reducing the NI part of the cross section in
that region. As stated before, without the inclusion 
of such a factor $C_{22}$, the over-prediction of the theoretical cross
section in the low $\nu$-region would be more pronounced.

Next, applying the prescription recalled in subsection~\ref{sec:is}, we make 
a comparison between GRS and IS cross sections for $E=3.595$ GeV. 
Fig.~5a (5b) shows the results for  $\theta=30^\circ$  using
the $\phi_0$ ($\phi_{01}$) approximation for the 
reduced response, together with the NE3 data. The agreement between the two
predictions is good, but not perfect. One of the causes is
undoubtedly the  use of (\ref{a87bis}), which is an approximation for
the parametrized, off-shell  total profile function $\tilde\Gamma$, and is of 
course not intrinsic to the actual ladder summation. Also the agreement 
with the  data is good, except for the smallest $\nu$.

One observes that the GRS and IS predictions using only $\phi_0$
diverge for decreasing $\nu$ and  that the GRS prediction is closest
to the data. This is shown in Fig.~5a for $\theta =30^{\circ}$,
$E=3.595$ GeV, but holds in fact for all  examined cases. Comparison
of Figs. (5a) and (5b) moreover shows that FSI for GRS are smaller
than for the IS, in particular for smaller $\nu$. The two observations 
above can be understood theoretically \cite{gr1} and have previously
been demonstrated for simple models. 

Finally,  one infers from Fig.~5b that the differences between the 0
order IS and GRS cross sections are noticeably reduced when the first
order FSI is included in both calculations. A similar comparison  is
shown in Figs.~5c and~5d for $\theta=45^\circ$ and again the agreement
is found to be  good  after the inclusion of the FSI.

The separate IS NE and NI parts for $E=3.595$ GeV and $\theta=30^\circ$
($\theta=45^\circ$) are shown in  Fig.~6a (Fig.~6b), where we used the same
notation adopted in Fig.~3. In this case, the NI part 
computed with the $\phi_0$ approximation for the reduced response
(thin solid line) stays well below the NE one (thin dashed line) in
the low $\nu$ region, as already found in Ref.~\cite{ciof1}.
Now, in the evaluation of the convolution~(\ref{a2}), $|y_0|$ becomes
rather large as $z\rightarrow 4$ and $\Fz$ rapidly decreases
($y_0\rightarrow -\sqrt{Q^2}$ as $z\gg1$). As a consequence, the IS integrand 
of Eq.~(\ref{a2}) is very small in the low $\nu$ region (in contrast
to what happens in the GRS case). However, as already  discussed
in relation with Fig.~5, there the FSI contributions in the IS case
are sizeable. In fact, now  the FSI are given by Eq.~(\ref{a88}) without 
the term $\tilde G^{(2)}_1$, which otherwise would partially cancel
the contribution of the large term $\tilde G^{(1)}_1$. 
As a result, in the low $\nu$ region, the IS NI cross section calculated
at the level of the 0+1 order becomes larger than the NE one, and
rather close to the GRS NI cross section.

Another interesting aspect is the influence of the  nuclear
interaction, chosen  to calculate the density matrices.
In Fig.~7 we display  for  $E=3.595$ GeV,
$\theta=30^\circ$  cross sections  computed on the basis of
the three previously mentioned models of nuclear interaction
(AV18UR, AV18 and AV14UR).  The calculations have been performed
using $\Fz$ from Eq.~(\ref{aa8b}) with the density matrices determined
directly from the corresponding nuclear wave functions. For identical
kinematics the predictions for 
AV18UR and AV14UR can hardly be distinguished, whereas the AV18 cross
section is slightly different from the other two in the low-$\nu$
tail: There clearly is only weak dependence on the nuclear interaction.

\section{Conclusions}
\label{sec:conc}

We have studied inclusive scattering of high-energy electrons from
${}^4$He, for energy losses
below and around the quasi-elastic peak and up into
the deep inelastic scattering region. The underlying  model assumes
non--interference between nucleonic and sub-nucleonic degrees of 
freedom, which implies that total
nuclear structures function may be expressed as a 
generalized convolution of the structure functions of free nucleons 
and the one  of a  nucleus composed of point particles. The model 
is estimated to become gradually imprecise for $Q^2\lesssim 2 - 2.5$ GeV$^2$. 
Structure functions for a nucleus of point-particles are computed via the 
reduced response $\phi(q,y)$ using  a  relativistic
generalization of the GRS series, which includes the   first order  FSI. 

A new element in the development  of the latter is an actual 
calculation of the required single- and two-particle, semi-diagonal density 
matrices, based on accurately computed  $^4$He ground state wave function. The 
above replaces previously used parametrizations of derived density matrices  
for targets with $A\ge$ 12. Computed cross sections appear to be hardly 
dependent on the choice of the $NN$ interaction.

We also exploited a prescription to derive from the GRS series to
order $1/q$ similar terms for the IS. The two methods produce  cross 
sections which  are rather similar, in particular after the
inclusion of the corresponding FSI contributions. 

In conclusion
our predictions are in good agreement with the NE3 
SLAC-Virginia data for all scattering
angles, in spite of the fact that for $\theta \lesssim 30^{\circ}$ the
involved $Q^2$ fall below the validity estimate.
We also computed cross sections  for a future  JLab experiment at
$E=6 $ GeV.  Its kinematics are largely within the estimated limits 
of the underlying theory and a comparison of theory with data will be
more significant for those than  is the case for the NE3 data.

\section{Acknowledgment}
 
The authors are grateful for having received from C. Ciofi degli Atti
tables of numerical values of the $^4$He spectral function, 
mentioned in Section II.  One of us (M.V.) also gratefully acknowledges a
helpful discussion with G.I. Lykasov.

\begin{figure}
\let\picnaturalsize=N
\def\picsize{5in}
\def\picfilenamea{fig3_6.eps}
\ifx\nopictures Y\else{\ifx\epsfloaded Y\else\input epsf \fi
\let\epsfloaded=Y
\centerline{
\ifx\picnaturalsize N\epsfxsize \picsize\fi \epsfbox{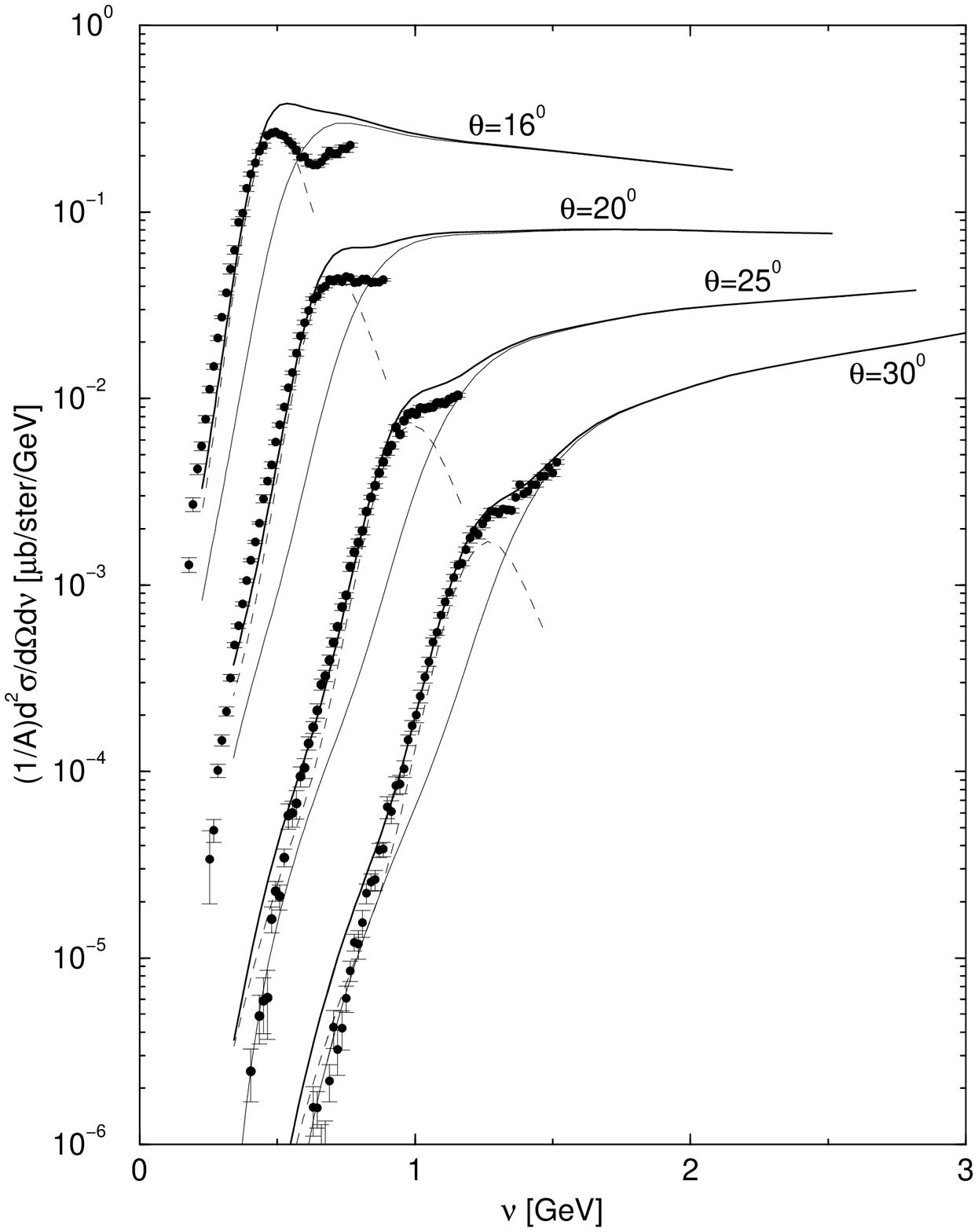}
 }}\fi
\caption{  
Predicted cross sections for inclusive scattering  of 3.595 GeV 
electrons from
$^4$He as function of the  energy loss $\nu$ and for four values of the
scattering angles $\theta$ in the (0+1)th order approximation
(thick solid lines).  Thin dashed (solid) lines show the NE (NI) part
of the cross sections. Data are from Ref.~\protect\cite{ne3}.
}
\end{figure}

\begin{figure}
\let\picnaturalsize=N
\def\picsize{5in}
\def\picfilenamea{fig6_0.eps}
\ifx\nopictures Y\else{\ifx\epsfloaded Y\else\input epsf \fi
\let\epsfloaded=Y
\centerline{
\ifx\picnaturalsize N\epsfxsize \picsize\fi \epsfbox{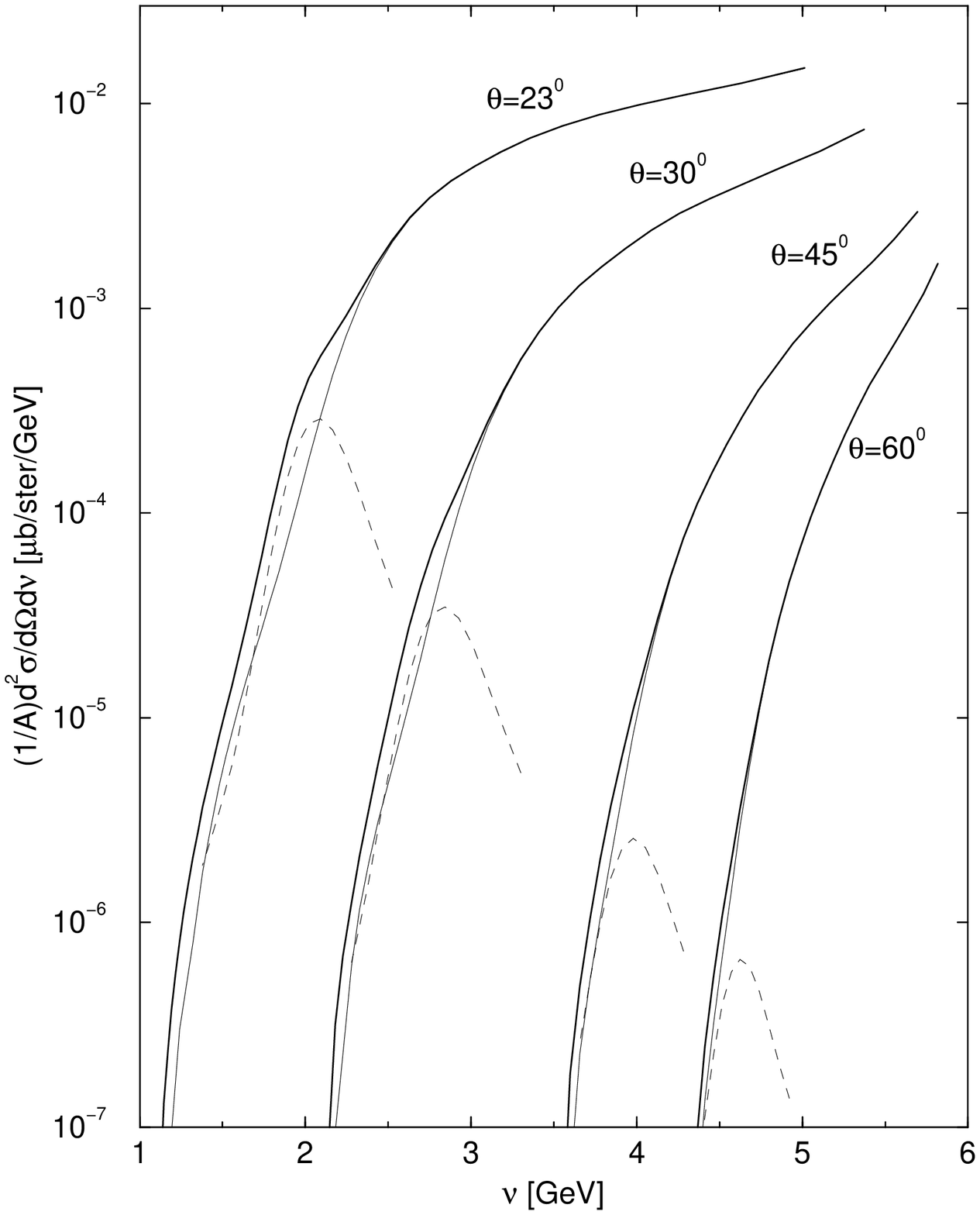}
 }}\fi
\caption{
Predicted cross sections  for inclusive scattering  of 6.0 GeV
electrons from
$^4$He as function of the  energy loss $\nu$ and for four values of the
scattering angles $\theta$ in the (0+1)th order approximation
(thick solid lines).  Thin dashed (solid) lines show the NE (NI) part
of the cross sections.
}
\end{figure}

\begin{figure}
\let\picnaturalsize=N
\def\picsize{6in}
\def\picfilenamea{fig_grs.eps}
\ifx\nopictures Y\else{\ifx\epsfloaded Y\else\input epsf \fi
\let\epsfloaded=Y
\centerline{
\ifx\picnaturalsize N\epsfxsize \picsize\fi \epsfbox{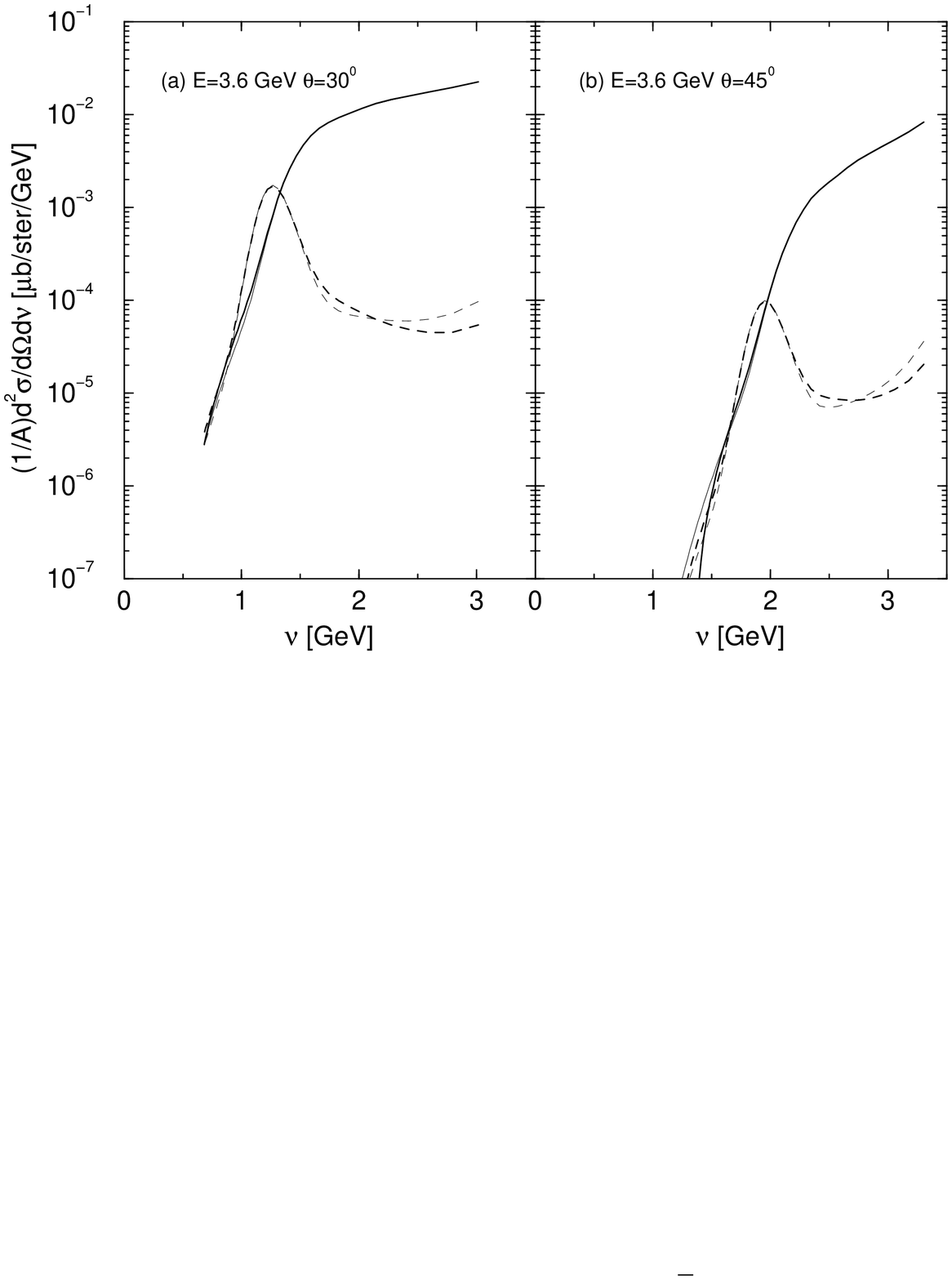}
 }}\fi
\caption{(a) Contributions to the cross section for $E=3.595$ GeV, 
$\theta=30^\circ$ from the NE (dashes) and the NI
(solid lines) parts of the nucleon SF. Thin and thick lines are
computed from the the 0th and (0+1)th order reduced response 
$\phi(\qq,y_G)$, respectively.  The density 
matrices correspond  to the  AV18UR nuclear interaction. (b) The same
as in (a), but for $\theta=45^\circ$.
} 
\end{figure}

\begin{figure}
\let\picnaturalsize=N
\def\picsize{4.5in}
\def\picfilenamea{figz.eps}
\ifx\nopictures Y\else{\ifx\epsfloaded Y\else\input epsf \fi
\let\epsfloaded=Y
\centerline{
\ifx\picnaturalsize N\epsfxsize \picsize\fi \epsfbox{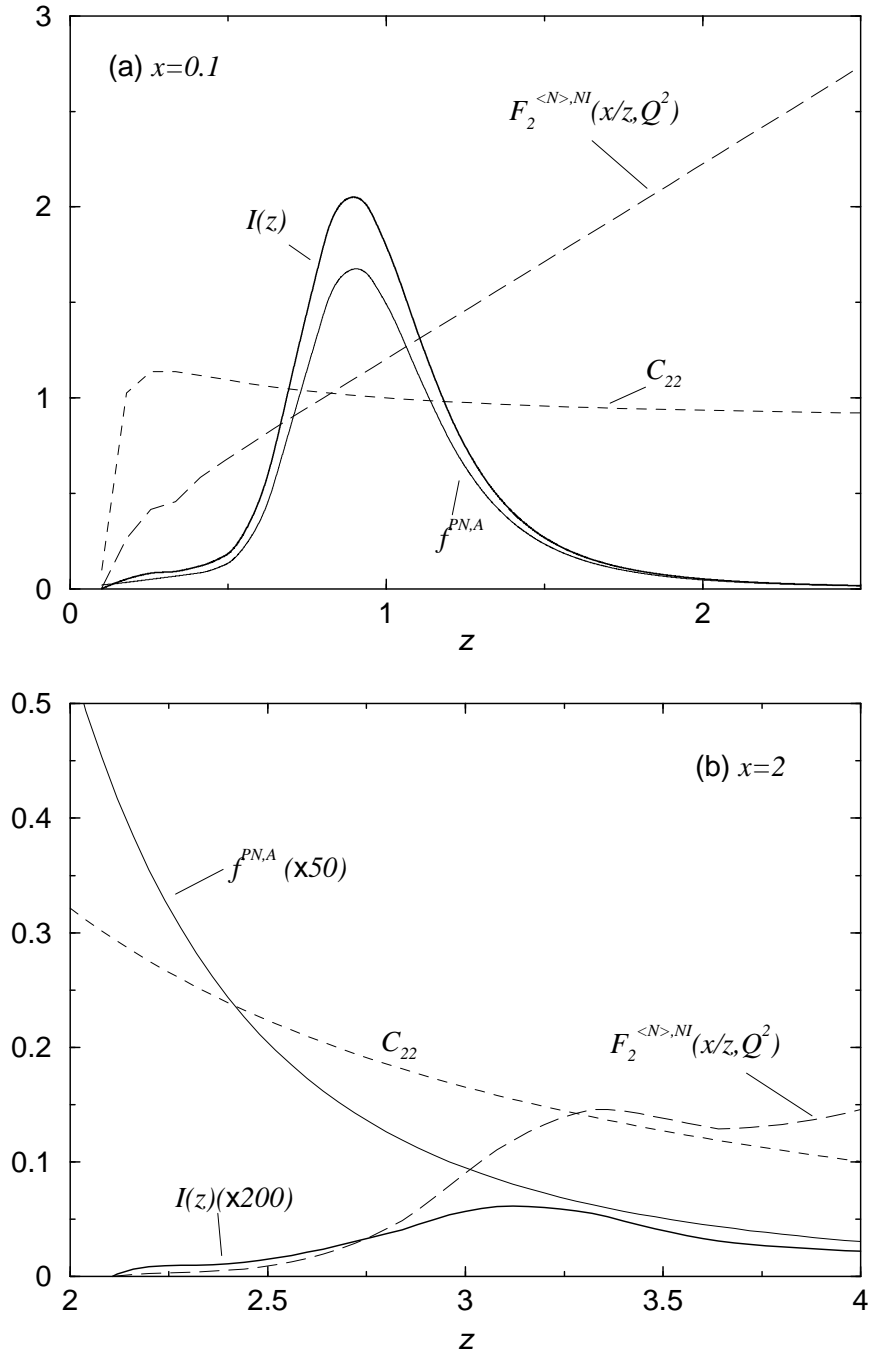}
 }}\fi
\caption{The GRS function $f^{PN,A}(z,Q^2)$ (thin solid line),
the nucleon SF $F^{\langle N\rangle, {\rm NI}}_2(x/z,Q^2)$ (long
dashed 
line) and the mixing factor $C_{22}(Q^2/x^2,z)$ (short dashed line) 
entering Eq.~(\ref{a2}) as function of $z$ for $E=3.595$ GeV,
$\theta=30^\circ$ 
and two cases of $x$: (a) $x=0.1$ (corresponding to have $Q^2=0.58$
GeV$^2$, $\nu=3.00$ GeV)  and (b) $x=2$ ($Q^2=2.76$ GeV$^2$,
$\nu=0.74$ GeV). $I(z)$ is the corresponding integrand of Eq.~(\ref{a2}).
} 
\end{figure}

\begin{figure}
\let\picnaturalsize=N
\def\picsize{6in}
\def\picfilenamea{fig_comp.eps}
\ifx\nopictures Y\else{\ifx\epsfloaded Y\else\input epsf \fi
\let\epsfloaded=Y
\centerline{
\ifx\picnaturalsize N\epsfxsize \picsize\fi \epsfbox{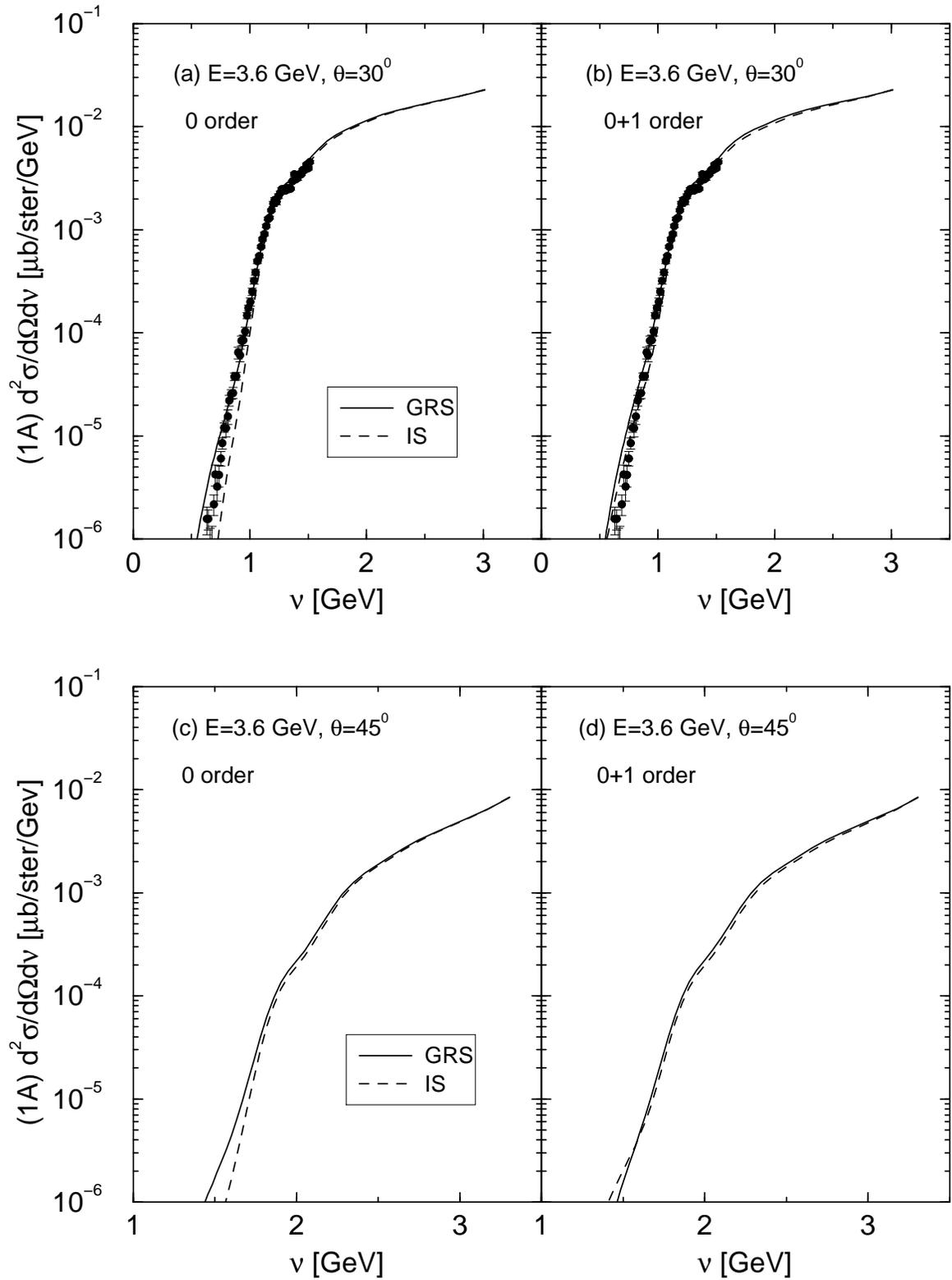}
 }}\fi
\caption{(a) Comparison of the 0th order GRS and derived IS cross
sections for $E=3.595$ GeV, $\theta=30^\circ$.  (b) As in (a) but in
the case of the (0+1)th order. (c) and (d): as in (a) and (b),
respectively, but for $\theta=45^\circ$. 
}
\end{figure}

\begin{figure}
\let\picnaturalsize=N
\def\picsize{6in}
\def\picfilenamea{fig_is.eps}
\ifx\nopictures Y\else{\ifx\epsfloaded Y\else\input epsf \fi
\let\epsfloaded=Y
\centerline{
\ifx\picnaturalsize N\epsfxsize \picsize\fi \epsfbox{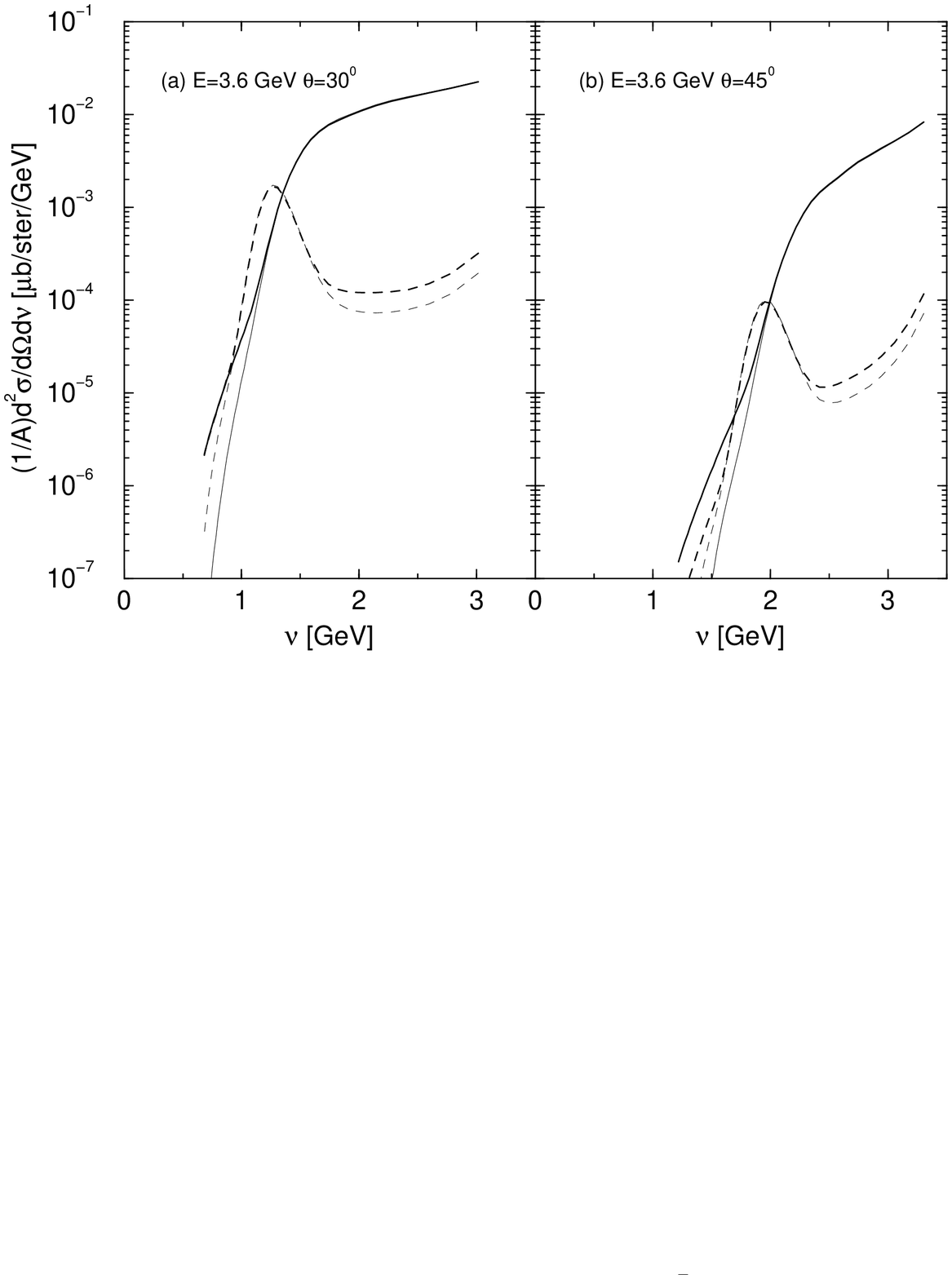}
 }}\fi
\caption{As in Fig.~3 but for the IS case. (a) Contributions to the
cross section for $E=3.595$ GeV,  
$\theta=30^\circ$ from the NE (dashes) and the NI
(solid lines) parts of the nucleon SF. Thin and thick lines are
computed from the the 0th and (0+1)th order reduced response 
$\phi(\qq,y_0)$, respectively.  The density 
matrices correspond  to the  AV18UR nuclear interaction. (b) The same
as in (a), but for $\theta=45^\circ$.
} 
\end{figure}

\begin{figure}
\let\picnaturalsize=N
\def\picsize{5in}
\def\picfilenamea{fignn.eps}
\ifx\nopictures Y\else{\ifx\epsfloaded Y\else\input epsf \fi
\let\epsfloaded=Y
\centerline{
\ifx\picnaturalsize N\epsfxsize \picsize\fi \epsfbox{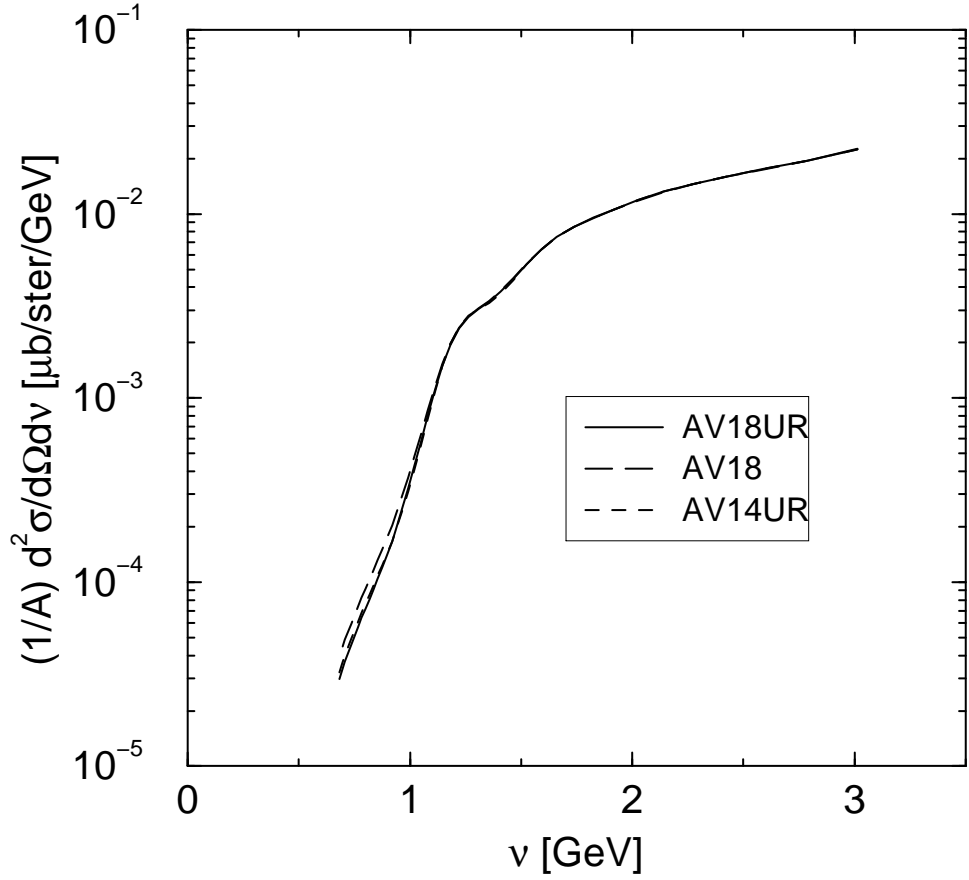}
 }}\fi
\caption{(0+1)th order GRS cross sections at $E=3.595$ GeV and
$\theta=30^\circ$ for the AV18 (dashed), AV14UR (long 
dashed) and AV18UR (solid) nuclear interactions.
}
\end{figure}

\begin{table}
\caption{ Selection of cross section ratios He/C, C/Fe. For given angle we
give $x,Q^2$ and the ratios. Former are from Ref. ~\protect\cite{ne3},
latter from Ref. \protect\cite{rtt}.
}
\begin{tabular}{ccc| ccc| ccc}
    \multicolumn{3}{c|}{$\theta=20^\circ$}
 &  \multicolumn{3}{c|}{$\theta=25^\circ$}
 &  \multicolumn{3}{c}{$\theta=30^\circ$}\\
\hline
       $x$ & $Q^2$  &He/C
      &$x$ & $Q^2$  &He/C
      &$x$ & $Q^2$  &He/C\\
\hline
       2.69  & 1.44 & 0.64
    &  2.10  & 2.07 & 0.64
    &  1.37  & 2.51 & 0.66  \\
       1.40  & 1.34 & 0.70
    &  1.23  & 1.88 & 0.69
    &  1.08  & 2.35 & 1.17  \\
       0.70  & 1.18 & 0.88
    &  1.37  & 2.51 & 0.96
    &  0.81  & 2.12 & 1.03  \\
\end{tabular}

\begin{tabular}{ccc| ccc}
   \multicolumn{3}{c|}{$\theta=15^\circ$}
 &  \multicolumn{3}{c}{$\theta=30^\circ$}\\
\hline
   $x$ & $Q^2$ &C/Fe
 & $x$ & $Q^2$ &C/Fe\\
\hline
   2.49  & 1.05 & 0.82
&  1.95  & 3.38 & 0.70 \\
   1.02  & 0.97 & 1.18
&  1.37  & 3.09 & 0.98 \\
   0.65  & 0.91 & 0.97
&  1.01  & 2.79 & 1.04 \\
   0.43  & 0.83 & 1.00
&  0.72  & 2.43 & 1.00 \\
\end{tabular}
\label{tab:rat}
\end{table}

\begin{table}
\caption{Binding energies in MeV of $^{4}$He calculated with the CHH method
using the AV18 and AV18/UIX, and the older AV14 and AV14/UVIII, 
Hamiltonian models.  Also listed are the corresponding 
\lq\lq exact\rq\rq GFMC results~\protect\cite{PPCPW97} 
as well as the experimental value.}
\begin{tabular}{l|c|c}
    Model    &  CHH & GFMC  \\
\hline
AV18     &  24.0  & 24.1(1)  \\
AV18/UIX   & 28.1 & 28.3(1)  \\
AV14     &  24.0  & 24.2(2)  \\
AV14/UVIII & 27.5 & 28.3(2)  \\
\hline
EXP&\multicolumn{2}{c}{28.3}
\end{tabular}
\label{tab:be}
\end{table}

\begin{table}
\caption{ Values of $\nu$ (in GeV) and $Q^2$ (in GeV$^2$) for
$E=3.6$ GeV~\protect\cite{ne3}.
The selected values of $x$ correspond approximately to the
QEP and the lower and upper ends of the curves shown in
Fig.~1.}
\begin{tabular}{c  |  c@{\qquad}c |  c@{\qquad}c |  c@{\qquad}c} 
 &  \multicolumn{2}{c}{lower end} 
 &  \multicolumn{2}{c}{QEP} 
 &  \multicolumn{2}{c}{upper end} \\
 &  \multicolumn{2}{c}{$x=2$} 
 &  \multicolumn{2}{c}{$x=1$} 
 &  \multicolumn{2}{c}{$x=0.1$} \\
\hline
$\theta$ & $\nu$ &  $Q^2$ 
         & $\nu$ &  $Q^2$ 
         & $\nu$ &  $Q^2$ \\
\hline
$16^\circ$ &  0.25  & 0.93 & 0.46  & 0.87 & 2.00  & 0.44 \\
$20^\circ$ &  0.37  & 1.40 & 0.68  & 1.27 & 2.50  & 0.48 \\
$25^\circ$ &  0.55  & 2.06 & 0.95  & 1.79 & 2.80  & 0.54 \\
$30^\circ$ &  0.74  & 2.76 & 1.22  & 2.29 & 3.00  & 0.58 \\
$45^\circ$ &  1.23  & 4.99 & 1.90  & 3.58 & 3.31  & 0.62 \\
\end{tabular}
\label{tab:ne3}
\end{table}

\begin{table}
\caption{ Values of $\nu$ (in GeV) and $Q^2$ (in GeV$^2$) for
$E=6.0$ GeV and the 
values of $\theta$ of the proposed CEBAF experiment.
The selected values of $x$ correspond approximately to the
QEP and the lower and upper ends of the curves shown in
Fig.~2.}
\begin{tabular}{c| c@{\qquad}c| c@{\qquad}c| c@{\qquad}c}
 &  \multicolumn{2}{c}{lower end} 
 &  \multicolumn{2}{c}{QEP} 
 &  \multicolumn{2}{c}{upper end} \\
 &  \multicolumn{2}{c}{$x=2$} 
 &  \multicolumn{2}{c}{$x=1$} 
 &  \multicolumn{2}{c}{$x=0.1$} \\
\hline
$\theta$ & $\nu$ &  $Q^2$ 
         & $\nu$ &  $Q^2$ 
         & $\nu$ &  $Q^2$ \\
\hline
$23^\circ$ &  1.21  & 4.56 &
              2.02  & 3.80 &
              5.01  & 0.94 \\
$30^\circ$ &  1.80  & 6.75 &
              2.77  & 5.20 &
              5.37  & 1.01 \\
$45^\circ$ &  2.90  & 10.90 &
              3.91  & 7.34 &
              5.70  & 1.07 \\
$60^\circ$ &  3.69  & 13.86 &
              4.57  & 8.58 &
              5.82  & 1.09 \\
\end{tabular}
\label{tab:arr}
\end{table}


\begin{references}
 
\bibitem{nik}
N.N. Nikolaev, J. Speth and B.G. Zakaharov, J. Exp. Theor. Phys.
82, 1046 (1996); A. Bianconi, S. Jeshonnek, N.N. Nikolaev
and B.G. Zakaharov, Phys. Lett. B 343, 13 (1995) 

\bibitem{rj}
A.S. Rinat, B. K. Jennings, Phys. Rev. C59, 3371 (1999)

\bibitem{ciof1}
See for instance: C. Ciofi degli Atti, E. Pace and G. Salm\`e, Phys. Rev. 
C 43, 1155 (1991);
C. Ciofi degli Atti, D.B. Day and S. Liuti, Phys. Lett. B 225, 215 (1989),
Phys. Rev. C46, 1045 (1992)

\bibitem{omar}
O. Benhar, A. Fabrocini, S. Fantoni, G.A. Miller, V.R. Pandharipande and
I. Sick, Phys. Rev. C44, 2328 (1991); Phys. Lett. B 359, 8 (1995)

\bibitem{cor}
P. Fernandez de Cordoba, E. Marco, H. Mutter, E. Oset and A. Faessler, Nucl.
Phys. A611, 514 (1996)

\bibitem{ciof3}
C. Ciofi degli Atti and S. Simula, Phys. Lett. B 325, 276 (1994)


\bibitem{grs}
H. Gersch, L.J. Rodriguez and Phil N. Smith, Phys. Rev. A5, 1547 (1973)

\bibitem{sag1}
S.A. Gurvitz, Phys. Rev. C42, 2653 (1990)

\bibitem{gr1} 
S.A. Gurvitz and A.S. Rinat, nucl-th/0106032; submitted to Phys. Rev. C.

\bibitem{rt1}
A.S. Rinat and M.F. Taragin, Nucl. Phys. A598, 349  (1996); $ibid$ A620,
412 (1997); Erratum $ibid$ A623, 773 (1997)

\bibitem{rtt}
A.S. Rinat and M.F. Taragin, Phys. Rev. C60, 044601 (1999)

\bibitem{rt2}
A.S. Rinat, Phys. Rev. B40, 6625 (1989); A.S. Rinat, M.F. Taragin, 
F. Mazzanti and A. Polls, Phys. Rev. B57, 5347 (1998)

\bibitem{MS91} 
H. Morita and T. Suzuki, Progr. Theor. Phys.  86, 671 (1991)

\bibitem{efrosS}
V.D. Efros, W. Leidemann and G. Orlandini, 
Phys. Rev. C 58, 582 (1998)

\bibitem{efros}
V.D. Efros, W. Leidemann and G. Orlandini, 
Phys. Lett. B 338, 130 (1994); Phys. Rev. Lett. 78, 432 (1997)

\bibitem{ne3}
D.B. Day $et\,al$, Phys. Rev. C48, 1849 (1993)

\bibitem{gr2}
S.A. Gurvitz and A.S. Rinat, TR-PR-93-77/ WIS-93/97/Oct-PH; Progress in
Nuclear and Particle Physics, Vol. 34, 245 (1995)

\bibitem{rt3}
A.S. Rinat, Proceedings of the Second International Conference on
'Perspectives in Hadronic Physics', ICTP, Trieste, Italy (1999),
S. Boffi {\it et al} Eds., World Scientific (Singapore), p.62

\bibitem{commar}
A.S. Rinat and M.F. Taragin, Phys. Rev. C62, 034602 (2000)

\bibitem{atw}
G.B. West, Ann. of Phys. (NY) 74, 646 (1972); W.B. Atwood and G.B. West,
Phys. Rev. D7, 773 (1973)

\bibitem{pion}
C.H. Llewellyn Smith, Phys. Lett. B 128, 117 (1983); M. Ericson and A.W.
Thomas, Phys. Lett. B 128, 120 (1983)
  
\bibitem{bod} 
A. Bodek and J.L. Ritchie, Phys. Rev.  D23, 1070 (1981)

\bibitem{amad}
P.  Amadrauz $et\,al$, Phys.  Lett.  B295, 159 (1992); M.  Arneodo $et\,
al\,,ibid$ B364, 107 (1995)

\bibitem{ciof4}
C. Ciofi degli Atti and S. Simula, Phys. Rev. C53, 1689 (1996)

\bibitem{rt4}
A.S. Rinat and M.F. Taragin, Nucl. Phys, A623, 519 (1997)

\bibitem{arr1} 
J. Arrington $et\,al$, Phys. Rev. C53, 2248 (1996)

\bibitem{arr}
J. Arrington $et\,al$, Phys. Rev. Lett. 82, 2056 (1999)

\bibitem{kam}
 H. Kamada $et\,al$, Phys. Rev. C64, 044001 (2001)

\bibitem{F83} M. Fabre de la Ripelle, Ann. Phys. (N.Y.)  147, 281        
(1983)

\bibitem{KRV95} M. Viviani, A. Kievsky and S. Rosati,  Few--Body Systems
                18, 25  (1995)

\bibitem{KRV94} A. Kievsky, S. Rosati, M. Viviani,
   Nucl. Phys.  A577,  511 (1994)

\bibitem{AV18} R.B. Wiringa, V.G.J. Stoks, and R. Schiavilla,
               Phys. Rev. C51, 38  (1995)

\bibitem{PPCPW97} B.S. Pudliner {\it et al.}, Phys. Rev.  C56,
                  1720 (1997)

\bibitem{AV14} R.B. Wiringa, R.A.  Smith, and T.L. Ainsworth,
               Phys. Rev.    C29, 1207 (1984)

\bibitem{W91} R. B. Wiringa, Phys. Rev.  C43, 1585 (1991)

\bibitem{arrd} J. Arrington, private communication; I. Niculescu $et \,al$, 
               Phys. Rev. Lett.    85, 1182 (2000)

\bibitem{rock}
S. Rock $et\,al$, Phys. Rev. 26, 1592 (1982)

\bibitem {arr2}
J. Arrington (spokesperson) {\it et al.}, ``A Precise Measurement of
the Nuclear Dependence of Structure Functions in Light Nuclei'',
Jefferson Lab. Proposal E-00-101, May 2000

\bibitem{lyka} See, for instance, O. Benhar, S. Fantoni and G. I. Lykasov,
               Eur. Phys. J. A5, 137 (1999) and references therein.
\end{references}
\end{document}